\documentclass[12pt,preprint]{aastex}
\usepackage{fullpage}
\usepackage[utf8x]{inputenc}
\usepackage{epsf,amsfonts,amsmath,amssymb}
\usepackage{graphicx}
\usepackage{times}
\usepackage{natbib}
\usepackage{ulem}
\usepackage{epsfig}
\usepackage{mathrsfs}
\usepackage{footmisc}
\usepackage[usenames,dvipsnames]{color}

\begin{document}  

\title{Shedding light on the prompt high efficiency paradox - self consistent modeling of GRB afterglows}
\author{Paz Beniamini\altaffilmark{1a}, Lara Nava\altaffilmark{1b}, Rodolfo Barniol Duran\altaffilmark{2c} \& Tsvi Piran\altaffilmark{1d}}
\altaffiltext{1}{Racah Institute for Physics, The Hebrew University, Jerusalem, 91904, Israel}
\altaffiltext{2}{Department of Physics and Astronomy, Purdue University, 525 Northwestern Avenue, West Lafayette, IN 47907, USA}
\email{(a) paz.beniamini@mail.huji.ac.il; (b) lara.nava@mail.huji.ac.il; (c) rbarniol@purdue.edu; (d) tsvi.piran@mail.huji.ac.il}
\begin{abstract}
We examine GRBs with both Fermi-LAT and X-ray afterglow data. Assuming that the $100\,$MeV (LAT)
emission is radiation from cooled electrons accelerated by external shocks, we show that the kinetic energy of the blast wave estimated from the $100\,$MeV flux is $\sim50$ times larger than the one estimated from the X-ray flux. 
This can be explained if either:  i) electrons radiating at X-rays are significantly cooled by SSC (suppressing the synchrotron flux above the cooling frequency) or ii) if the X-ray emitting electrons, unlike those emitting at $100\,$MeV energies,
are in the slow cooling regime. In both cases the X-ray flux is no longer an immediate proxy of the blast wave kinetic energy.
We model the LAT, X-ray and optical data and show that in general these possibilities are consistent with the data, and explain the apparent disagreement between X-ray and LAT observations.
All possible solutions require weak magnetic fields: $10^{-6}\lesssim \epsilon_B \lesssim 10^{-3}$ (where $\epsilon_B$ is the fraction of shocked plasma energy in magnetic fields).  
Using the LAT emission as a proxy for the blast wave kinetic energy we find that 
the derived prompt efficiencies are $\sim15\%$. This is considerably lower compared with previous estimates (87$\%$ and higher for the same bursts).
This provides at least a partial solution to the ``prompt high efficiency paradox”.
\end{abstract}

\section{Introduction}
\label{Int}
Afterglow theory predicts that synchrotron flux at frequencies larger than $\nu_c$ and $\nu_m$
($\nu_c$ is the cooling frequency and $\nu_m$ is the typical synchrotron frequency) is determined mostly by the kinetic energy of the blast wave ($E_{kin}$)
and the fraction of energy in shock-accelerated electrons ($\epsilon_e$). Importantly, this flux does not depend on the GRB environment and it depends
very weakly on the fraction of energy in the magnetic field, $\epsilon_B$ and the power law index of the electrons' energy distribution, $p$
(\citealt{Kumar(2000),Freedman(2001)} or, more recently, \citealt{Nava(2014)}).
For $p$ ranging from 2.1 to 2.5 the predicted flux changes by a factor of order unity.
Assuming a typical $\epsilon_e \approx0.1$ one can derive an estimate of the kinetic energy of the blast-wave
from afterglow observations at frequencies above $\nu_c$.
Comparing this energy with the energy radiated at the prompt phase ($E_{\gamma}$), yields an estimate of the prompt efficiency: $\epsilon_{\gamma}=E_{\gamma}/(E_{kin}+E_{\gamma})$.

This method has been often applied to pre-Swift X-ray observations at $\sim $day to estimate the kinetic energy of GRBs. It suggested that the efficiency of the prompt phase of GRBs should be:
$\epsilon_{\gamma}>0.5$ \citep{Frail(2001),Panaitescu(2001a),Panaitescu(2001b),Berger(2003)}.
The discovery of the X-ray plateaus in many of the Swift GRBs, led to an increase in the severity of the efficiency problem. The X-ray flux
at the beginning of the plateau phase (around 500 sec) is lower by $\sim 3$ as compared with the same flux estimated by extrapolating backwards in time the observations at $\sim$day and
therefore leads to an estimate of the kinetic energy lower by the same factor and to efficiencies of up to 0.9 \citep{Granot(2006),Fan(2006),Ioka(2006),Nousek(2006),Zhang(2006),Nysewander(2009)}.
Such high efficiencies pose an immense theoretical difficulty for prompt emission models.
This efficiency includes two distinct processes, the conversion efficiency of bulk to internal energy, i.e. the
efficiency of the energy dissipation process, and the conversion efficiency from internal energy
to radiation. Therefore, a high overall efficiency implies high efficiencies for both processes, which is not at all trivial.

A crucial point in this analysis is the assumption that at around $1\,$day the X-ray band is above the cooling frequency 
and that the flux at this frequency is not affected by SSC losses, as was pointed out by \cite{Fan(2006)}.
We examine here whether the X-ray flux is indeed a good proxy for the kinetic energy.
To do so, we need to pin down the location of $\nu_c$ at late times, assess the role of SSC cooling,
and re-derive the efficiency in case the basic assumptions (i.e., negligible Compton cooling and $\nu_X>\nu_c$ at the time of observation) are not valid.
As the position of $\nu_c$ and the Compton parameter, depend on $\epsilon_B$ and $n$ this will require us to self-consistently determine these parameters.
To do this, we collect LAT, X-ray and optical data (when available) for a sample of LAT GRBs. We assume that LAT photons are produced by external shocks (via synchrotron or SSC)
and find (both analytically and numerically) self-consistent solutions for the given set of observations. 

\section{The Sample}
\label{Sample}
We have collected a sample of GRBs detected both by Fermi-LAT and by Swift-XRT.
Since we are interested in cases in which both the LAT and the XRT emissions are most likely afterglow radiation from external shocks,
we included in our sample only those bursts for which the LAT emission lasted longer than the prompt phase.
We also consider optical observations, when available.
The final sample includes nine GRBs: 080916C, 090323, 090328, 090510, 090926A, 100414A, 110625A, 110731A and 130427A.

\section{The prompt efficiency - an apparent inconsistency}
\label{efficiency}
Our basic assumption is that LAT photons are produced by synchrotron emission from the forward shock. In this case we
can calculate the kinetic energy of the blast wave both from X-ray and LAT data by inverting the equation $F_\nu(\nu>\nu_c)\propto\epsilon_e^{p-1} E_{kin}^{2+p \over 4} \nu^{-p/2}t^{2-3p \over 4}$.
We find that the kinetic energies inferred from the LAT flux are on average $\sim 50$ times larger than those derived from X-rays.
In turn, the efficiency derived from LAT is much smaller (0.15 on average as opposed to 0.87 obtained from X-rays).

There could be two solutions to the apparent discrepancy between the energies derived from LAT and X-rays. First, the X-ray flux could be suppressed due to IC losses. 
If Compton losses are important, than a factor $(1+Y)^{-1}$ (where $Y$ is the Compton parameter) should be added to the equation for the flux above the cooling frequency: $F_{\nu}(\nu>\nu_c)\propto E_{kin}^{(2+p)/4}(1+Y)^{-1}$,
and therefore $\frac{E_{kin,X}}{E_{kin,LAT}} \propto (\frac{1+Y_X}{1+Y_{LAT}})^{4/(2+p)}$ ($Y_X$ is the Compton
parameter for X-ray radiating electrons and $Y_{LAT}$ is that of electrons radiating at the LAT band).
Since LAT photons are typically above the Klein-Nishina (KN) frequency, IC scattering of these photons is highly suppressed, and as a result $Y_{LAT}\lesssim1$.
However, X-ray emitting electrons are not in the KN regime therefore their SSC losses can explain the discrepancy in flux.
We refer to this possibility as ``SSC-suppression". A second possibility is that the X-ray band is actually below the cooling frequency.
In this spectral regime the flux depends strongly also on $\epsilon_B$ and on the external density. We refer to this possibility as ``slow cooling" scenario.
In both cases the X-ray flux is no longer a good proxy for the energy.
If one of the two scenarios can account for the data, then the kinetic energy estimated by the LAT flux is a better
estimate of the kinetic energies.

\section{Numerical results}
\label{numres}
We turn now to a general numerical solution, assuming only that both the X-ray and LAT photons are produced by the same external shock.
We relax the simplifying assumptions made above\footnote{Except for the assumption $\epsilon_e=0.1$.}, such as the assumption that the cooling frequency
is below the LAT band at the time of LAT observations or even that the LAT and X-ray fluxes are dominated by synchrotron and not SSC. For each burst,
we run over all possible values of $\epsilon_B, n, E_{kin}$ and change $p$ in the range: 2.1 to 2.8. For each set of values we use \cite{Nakar(2009)} to calculate
the Compton parameter $Y(\nu)$, including possible corrections due to KN effects, and estimate the synchrotron SED (using \citealt{Granot(2002)} with the addition of the effect of SSC suppression)
and the IC SED (using \citealt{Nakar(2009)}). We then calculate the fluxes at: $t_{LAT},t_X, t_{opt}$ (the latter is the time of optical observations, in case such observations are available)
at  $100\,$MeV, keV, eV respectively. We compare these fluxes with the observed values, and in case the difference is within the reported uncertainty of the observations and the accuracy of the model
(which we estimate to be up to  $\sim50\%$), we accept the solution.

\begin{figure}
\includegraphics[scale=0.3]{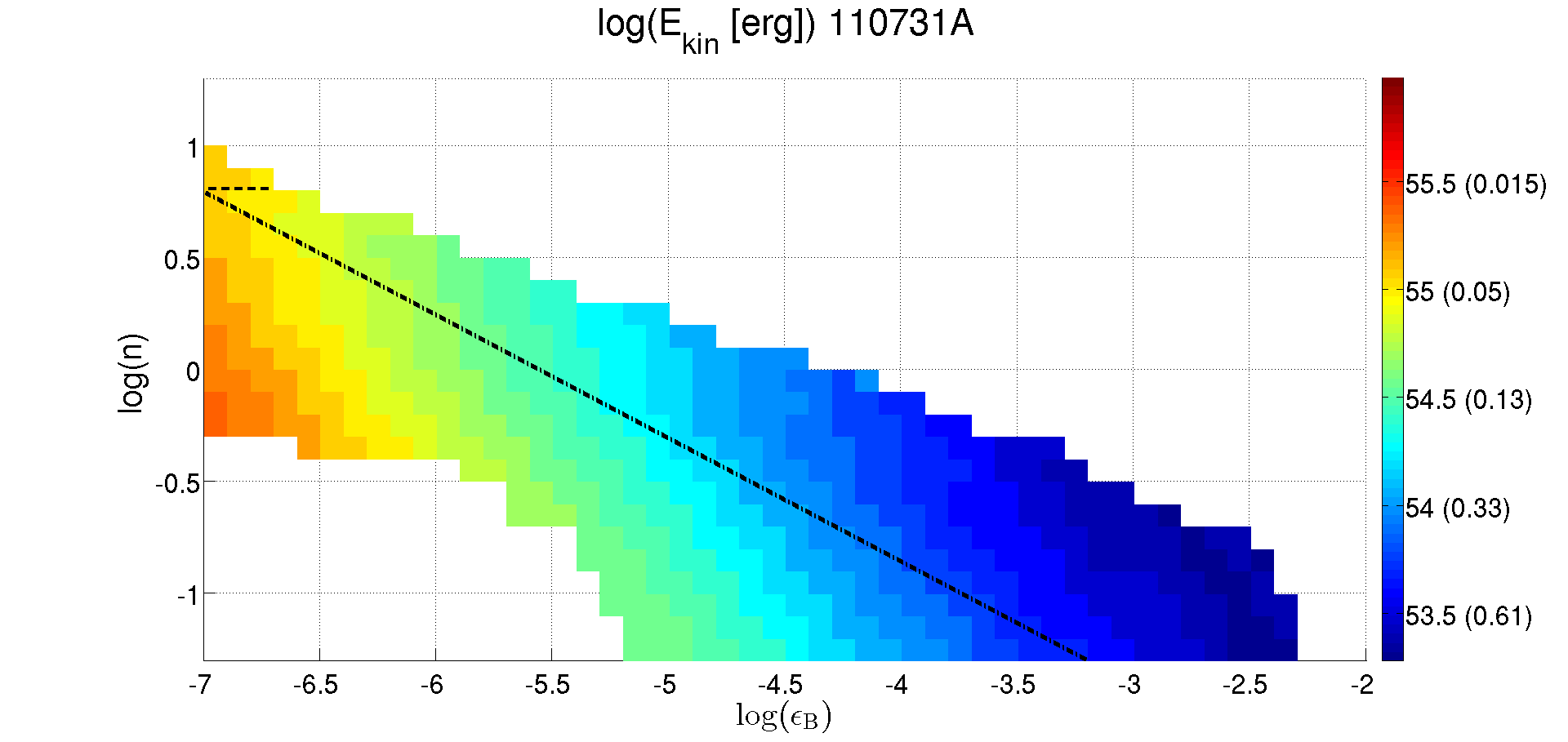}
\caption{The allowed parameter space for GRB 110731A in an ISM environment. Colours depict isotropic equivalent kinetic energies (or equivalently, values for the prompt efficiency in parenthesis)
for each point in the ($\epsilon_B, n$) plane.
The lower branch (below the dot-dashed line) corresponds to the ``slow cooling" solutions. Solutions above this line correspond to ``SSC suppressed" solutions.
The uppermost branch (in the upper left corner, above the dashed line) corresponds to situations in which X-rays are synchrotron from SSC-suppressed fast cooling electrons and the LAT is dominated by SSC photons.
However these solutions are ruled out by more detailed modeling of the spectrum.
For reasonable values of $3\times10^{-2}\mbox{  cm}^{-3} \lesssim n\lesssim 30 \mbox{ cm}^{-3} $, we obtain:  $\epsilon_{\gamma}\lesssim 0.6$, $E_{kin}\gtrsim 5 \times 10^{53}$ and $\epsilon_B \lesssim5\times 10^{-3}$
independent of the type of solution.}
\label{fig:110731A}
\end{figure}

We find a good agreement between the numerical calculations and the simplified analysis outlined above.
A typical example of allowed parameters space for GRB 110731A in an ISM environment can be seen in Fig. \ref{fig:110731A}. Three types of solutions can be seen in this figure. The lower branch
(lower densities, below the dot-dashed line) branch consists of solutions where X-rays are synchrotron radiation from slow cooling electrons, whereas the LAT flux is produced by synchrotron fast cooling electrons.
This is the slow cooling case in \S \ref{efficiency}. The intermediate branch (between the dot-dashed and the dashed line), consists of solutions where both X-rays and LAT are dominated by fast cooling synchrotron,
but the X-ray emitting electrons are suppressed by SSC. This corresponds to the SSC suppressed case discussed in \S \ref{efficiency}. The uppermost branch of solutions (above the dashed line),
consists of cases where the X-rays are still produced by synchrotron from SSC-suppressed fast cooling electrons, but the LAT photons are dominated by the SSC component instead of synchrotron.
In this case the spectral slope of $\nu F_{\nu}$ in LAT is rising ($\sim 0.5$) and is therefore in contradiction with LAT observations, ruling out this possibility
Both allowed solutions suggest that the prompt efficiency is moderate: $\epsilon_{\gamma}\leq 0.6$, corresponding to isotropic equivalent kinetic energies satisfying: $E_{kin}\geq 3\times 10^{53}$erg.
In addition, the allowed parameter space results in upper limits on the magnetization: $\epsilon_B \leq 5\times10^{-3}$.

\begin{figure}
\includegraphics[scale=0.4]{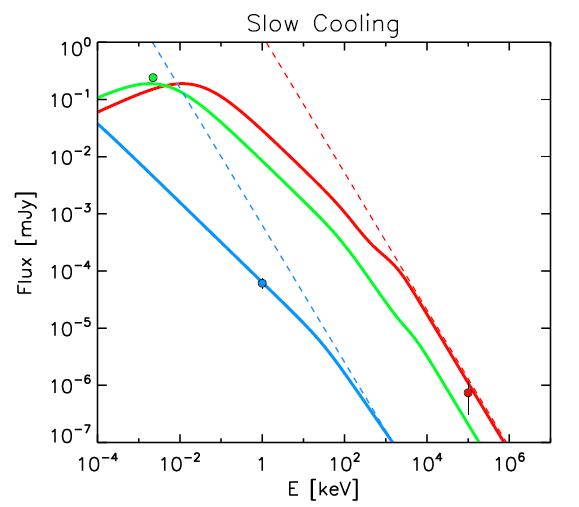}
\includegraphics[scale=0.4]{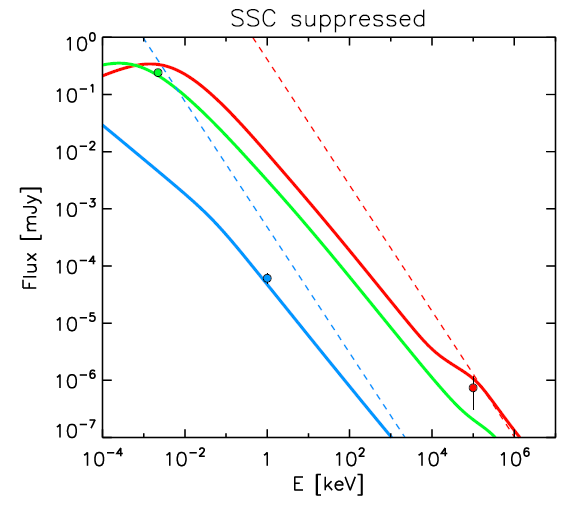}
\caption{Observed and predicted spectra for GRB 110731A. Left: ``slow cooling" SED for $\epsilon_B=8\times10^{-5}, n=0.015\mbox{ cm}^{-3}, E_{kin}=2\times 10^{54} erg, p=2.4$ (corresponding to a typical ``slow cooling'' type solution).
Right: ``SSC suppressed" SED for $\epsilon_B=2.6\times10^{-5}, n=0.7\mbox{ cm}^{-3}, E_{kin}=10^{54} erg, p=2.2$ (corresponding to a typical ``SSC suppressed'' type solution).
The solid curves are the results of the simulation. The red curve is the SED at the time of LAT observation (at 300 sec), the blue curve is the SED at the time of X-ray observations
(at 1.22 days) and the green curve is the SED at the time of optical observations (at 0.012 days). The observations in these three bands are denoted by filled circles with error-bars.
The red and blue dashed lines are the LAT and X-ray flux (accordingly) assuming fast cooling synchrotron with no IC suppression. It can readily be seen that there is no significant suppression
for LAT flux whereas for X-rays the suppression is by a factor of $\sim 10$ as compared with fast cooling synchrotron with no IC suppression.}
\label{fig:SED}
\end{figure}

Typical SEDs of the ``slow cooling'' and ``SSC suppressed'' solutions,  presented in Fig. \ref{fig:110731A} can be seen in Fig. \ref{fig:SED}. As shown in the figures, the X-ray flux has to be significantly suppressed as compared
with the flux obtained by assuming fast cooling synchrotron in this band. The LAT flux is typically not affected by IC suppression.

\section{Conclusions}
We have examined a sample of GRBs with both LAT and X-ray afterglow data.
Our basic assumption is that LAT photons (or at least the late tail of these photons) are produced by synchrotron emission from the forward shock.
We find that the X-ray flux is not a good proxy for the kinetic energy of the blast wave, either because the X-ray emitting electrons are slow cooling
or because the X-ray flux is suppressed by SSC losses. The LAT flux, is not prone to these problems and is therefore a better proxy of the energy.
The derived prompt efficiencies using LAT photons are considerably lower as compared with efficiency estimates using the X-rays
(15$\%$ as opposed to 87$\%$ and higher). 
These afterglow models require small values of $\epsilon_B$:  $10^{-6}\lesssim \epsilon_B \lesssim 10^{-3}$.
Such low values would arise if either there is no extra amplification of the external magnetic field beyond shock compression (see \citealt{RBD(2014)}) or if
strong magnetic fields are created by micro-turbulence near the shock, but then these fields decay rapidly with the distance from the shock front \citep{Derishev(2007),Lemoine(2013)}
and the bulk of the afterglow emission is produced in a low magnetic field region.

This work was supported by a Marie Curie IEF (LN), by the ERC grant GRBs, by a grant from the Israel ISF - China NSF collaboration,
by a grant from the Israel Space Agency, and by the I-Core Center of Excellence in Astrophysics.

\appendix
\numberwithin{equation}{section}

\end{document}